# V-SQL: A View-based Two-stage Text-to-SQL Framework

Zeshun You

Artificial Intelligence Department; Guangzhou Huaweisoft Co. Ltd; Hanjin Str. 1, 510635, Guangzhou, China; phone: +86 18520620934; e-mails: youzeshun@huaweisoft.com

Jiebin Yao

Artificial Intelligence Department; Guangzhou Huaweisoft Co. Ltd; Hanjin Str. 1, 510635, Guangzhou, China; phone: +86 18023757075; e-mails: yaojiebin@huaweisoft.com

Dong Cheng

Artificial Intelligence Department; Guangzhou Huaweisoft Co. Ltd; Hanjin Str. 1, 510635, Guangzhou, China; phone: +86 18565542915; e-mails: chengd@huaweisoft.com

Zhiwei Wen

Artificial Intelligence Department; Guangzhou Huaweisoft Co. Ltd; Hanjin Str. 1, 510635, Guangzhou, China; phone: +86 15768613940; e-mails: wenzw@huaweisoft.com

Zhiliang Lu

Artificial Intelligence Department; Guangzhou Huaweisoft Co. Ltd; Hanjin Str. 1, 510635, Guangzhou, China; phone: +86 13560017067; e-mails: luzhiliang@huaweisoft.com

Xianyi Shen

Artificial Intelligence Department; Guangzhou Huaweisoft Co. Ltd; Hanjin Str. 1, 510635, Guangzhou, China; phone: +86 18665731100; e-mails: shenxy@huaweisoft.com

Corresponding author: shenxy@huaweisoft.com

The text-to-SQL task aims to convert natural language into Structured Query Language (SQL) without bias. Recently, text-to-SQL methods based on large language models (LLMs) have garnered significant attention. The core of mainstream text-to-SQL frameworks is schema linking, which aligns user queries with relevant tables and columns in the database. Previous methods focused on schema linking while neglecting to enhance LLMs' understanding of database schema. The complex coupling relationships between tables in the database constrain the SQL generation capabilities of LLMs. To tackle this issue, this paper proposes a simple yet effective strategy called view-based schema. This strategy aids LLMs in understanding the database schema by decoupling tightly coupled tables into low-coupling views. We then introduce V-SQL, a view-based two-stage text-to-SQL framework.  V-SQL involves the view-based schema strategy to enhance LLMs' understanding of database schema. Results on the authoritative datasets Bird indicate that V-SQL achieves competitive performance compared to existing state-of-the-art methods.

## 1. Introduction

The text-to-SQL task is to convert natural language queries into executable SQL statements, which facilitates access to databases for non-technical users. text-to-SQL has been utilized in robotic navigation, customer service and question answering.

In recent years, with the advancement of Large Language Models (LLMs), LLM-based text-to-SQL has become an active area of research. By utilizing their advanced language understanding capabilities, LLMs have demonstrated notable performance in text-to-SQL tasks even without fine-tuning. It is quite challenging for LLMs to accurately generate the final SQL in a single turn.

some prior works had divided text-to-SQL task into two stages. The first stage is schema linking, which aims to align the user's query with the relevant tables and columns in the database.The second stage is Logical Synthesis, designed to response user by generating SQL based on the user's query and the relevant table schema. As the key component of the text-to-SQL task, schema linking directly influences the effectiveness of systems. However, the hallucination problem in LLMs severely impacts the performance of schema linking. The hallucination is related to the complexity of database design patterns. The table structure design of relational database is primarily focused on the efficiency of maintaining data consistency, causing LLMs are difficult to accurately utilize database information. For example, it is unreliable that LLMs match tables and columns while use "select" or "where" in the SQLs containing join operations. As the red font shown in Table 1, the erroneous SQL mistakenly treats the column 'attribute_name' as a column under the table 'hero_attribute' and causes exception "no such column: hero_attribute.attribute_name.". The above phenomenon is referred to as schema-based hallucinations.

We assume that the cause of LLMs's hallucinations is foreign key relationships between tables are not well understood by LLMs.

SQL generation accuracy on simpler text-to-SQL database schemas shows higher performance compared to more complex database schemas. For example, performance on the WikiSQL dataset surpasses that on the Bird dataset. Based on these findings, a strategy is proposed by this paper to reduce LLM hallucinations through schema simplification using views, termed a view-based schema. This strategy map multiple related tables in the database into a single view and eliminate foreign key dependencies between the tables. The view-based schema strategy mitigate hallucinations by simplifying the database schema to reduce the number of complex join operations.

Furthermore, we propose V-SQL, a two-stage text-to-SQL framework. In the first stage, dummy SQL is generated based on the view-based schema strategy. In the second stage, this dummy SQL is reconstructed into final SQL to avoid errors introduced by the join operation of the view creation.

Our main contributions can be summarized as follows:

1. We introduced a simply and effective strategy for mitigate hallucinations, named view-based schema. This strategy creates loosely coupled views through table mapping, replacing tables for SQL generation.

2. Based on the view-based schema strategy, we proposed a novel text-to-SQL framework, V-SQL, and compared it with three baseline models on the Bird datasets. Experimental results demonstrate that this framework achieves competitive performance.

## 2. RELATED WORKS

### 2.1 LLM-based text-to-SQL

The implementation of existing LLM-based applications primarily depends on in-context learning (ICL) through prompt engineering[1]. ICL is a paradigm that enables language models to perform tasks using only a few examples in the form of demonstrations, or even without any examples. It does not require additional training and can be directly applied to pretrained LLMs[2].

**Table 1**

An example of a complex foreign key relationship leading to erroneous SQL generated by LLMs.

| Query | Please list all the superpowers of 3-D Man. |
|---|---|
| Goal SQL | SELECT COUNT(`superhero`.`id`) <br> FROM `superhero` <br> INNER JOIN `hero_attribute` ON `superhero`.`id` = `hero_attribute`.`hero_id` <br> INNER JOIN `attribute` ON `hero_attribute`.`attribute_id` = `attribute`.`id` <br> INNER JOIN `gender` ON `superhero`.`gender_id` = `gender`.`id` <br> WHERE `attribute`.`attribute_name` = 'Strength' <br> AND `hero_attribute`.`attribute_value` = 100 <br> AND `gender`.`gender` = 'Female' |
| Erroneous SQL | SELECT COUNT(`superhero`.`id`) FROM `superhero` <br> INNER JOIN `hero_attribute` ON `superhero`.`id` = `hero_attribute`.`hero_id` <br> INNER JOIN `attribute` ON `hero_attribute`.`attribute_id` = `attribute`.`id` <br> INNER JOIN `gender` ON `superhero`.`gender_id` = `gender`.`id` <br> WHERE `hero_attribute`.`attribute_name` = 'Strength' <br> AND `hero_attribute`.`attribute_value` = 100 <br> AND `gender`.`gender` = 'Female' |

### 2.2 LLM-based two-stage text-to-SQL

Two-stage reasoning framework is a typical LLM-based approach, which decomposing a complex SQL generation task into two simpler steps: specifically, 1) schema linking and 2) logical synthesis. Previous work often treats schema linking as a key optimization point, focusing on accurately linking user queries to the corresponding fields in the database. For example, in [3], a table schema classifier is developed to predicts the relevance scores of tables and columns based on user queries. Using these scores, they retain the table schema most relevant to the user's questions. This approach requires additional training of a model. Since the trained model lacks the strong domain transfer capabilities of larger models. TA-SQL[2] promotes incremental generalization by first generating a dummy SQL query and subsequently extracting relevant schema entities from it as the final output. This process helps in reasoning through the transformation logic in the subsequent stage. In [4], LLMs are instructed to generate a preliminary SQL query (PreSQL), then the table and column entities mentioned are identified as the linking results. However, previous two-stage SQL generation methods based on LLMs overlooked the challenges posed by complex foreign key relationships between tables in database schemas. Queries involving multiple tables often result in errors due to difficulties in comprehending their relationships.

## 3. Preliminaries

Given a natural language question $Q=\{q_1,\cdots,q_{|Q|}\}$, and its corresponding database schema $D=(C,T)$, where $C=\{c_1,\cdots,c_{|C|}\}$ represents the columns and $T = \{t_1,\ldots, t_{|T|}\}$ represents the tables. The objective of text-to-SQL is to generate the corresponding SQL query $y$. The LLM-based text-to-SQL process for generating an executable SQL query $y$ can be expressed as follows:

$$y = f(x, I, S, D \mid \theta), \quad (1)$$

Where the mapping function $f(\cdot|\theta)$ is applied by the LLM. The model's parameter $\theta$ is frozen during the ICL-based text-to-SQL task. $I$ denote the instruction for the text-to-SQL task, serving as a guiding to prompt the LLMs in producing a precise SQL query. $S$ is a set of few-shot input-output prompting demonstrations $\{(x_1, y_1),\cdots, (x_k, y_k)\}$. When $S$ is absent, the scenario becomes zero-shot prompting, where the model must rely solely on its pre-trained knowledge to generate SQL query. The result of this zero-shot prompting can be represented as:

$$y = f(x, I, D \mid \theta), \quad (2)$$

**Figure 1**

An illustration of V-SQL, utilizing the view-based schema strategy to map tables into views (A), generating dummy SQL based on the view (B), and reconstructing the dummy SQL into final SQL(C).

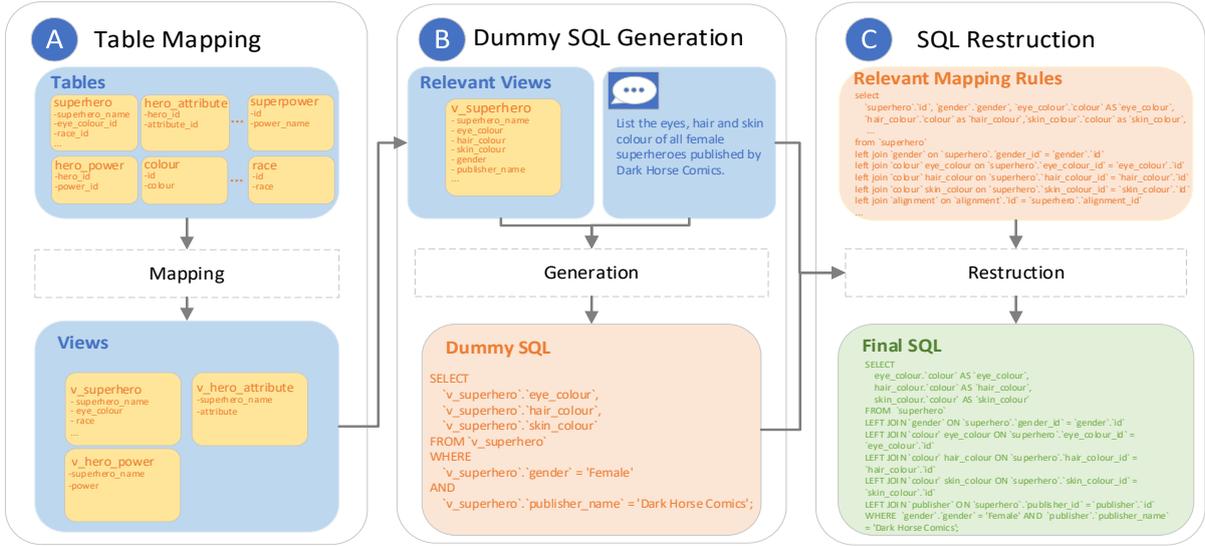

## 4. METHOD

**4.1 View-based schema**

Inspired by the phenomenon that simpler table structures are easier for LLMs to process than complex ones, we introduce a view-based schema strategy to mitigate hallucinations in text-to-SQL task. More precisely, in relational database design, foreign keys link multiple tables to maintain data consistency and relationships. However, this increases the complexity of the database schema, making it more challenging for LLMs to accurately interpret the schema, which can result in hallucinations.

A view is a virtual table in a relational database that does not store data but represents a SQL query. The SQL query is executed when the view is accessed. The view-based schema strategy simplifies the database schema by mapping related columns distributed across multiple tables into a single view. As shown in Figure 1(A), the eye_colour_id in the superhero table can be joined with the colour table to obtain the eye_colour column in the new view v_superhero. The process of generating view-based SQL is represented by the following formula:

$$y = f(x, I, S, D' \mid \theta), \qquad (3)$$

Where $D'$ denotes the database schema composed of views. It is used to replace $D$ in providing the database information to the LLM.

**4.2 V-SQL**

We further propose a text-to-SQL framework called V-SQL, based on the view-based schema strategy. V-SQL draws inspiration from the autoencoder architecture and consists of two stages. The first stage is akin to an encoder, encoding the user query into a dummy SQL based on views with columns mapped from multiple tables. The dummy SQL is akin to a high-dimensional vector output by encoder of autoencoder, which is shorter than final SQL and has a higher information density as shown by the orange box in Figure 1(B). The second stage is akin to decoder, reconstructing the dummy SQL into final SQL using the mapping rule for views creation, as illustrated in Figure 1(C).

V-SQL is a concise and efficient text-to-SQL framework that mitigates hallucinations and reduces token usage. The three modules of V-SQL, as shown in Figure 1, are detailed in the following sections. The prompts used in each module are provided in the Appendix.

**4.3 Table Mapping module**

Given a natural language question Q and the corresponding database schema D, previous approaches first perform schema linking to roughly identify which tables and columns are relevant to Q. However, large relational databases contain numerous foreign keys between tables, which increases the complexity of the database schema. Due to the complex database schema $D$, schema linking becomes a bottleneck in previous text-to-SQL framework. To address this challenge, we transform D into highly independent views, allowing the LLM to generate accurate SQL with fewer join operations.

As shown in Figure 2 of Appendix, we designed a prompt to assist in creating table mappings. This prompt allows LLM to generates the syntax for view creation, interpreting *D* as *D'*. *D'* serves as part of the prompt for LLM in generating and reconstructing the dummy SQL, as shown in Figures 1(B) and (C). It should be noted that the database schema *D'* function as guide for LLM and does not require creation in database.

### 4.4 Dummy SQL Generation

The dummy SQL serves as an intermediate result, generated by the LLM based on user query Q and its corresponding view. The dummy SQL is not executed to avoid errors introduced by the join method. Specifically, a view represents a query result. Querying the view involves further querying within an existing result set. Therefore, performing a query on a view will be influenced by the query results that created the view, which may introduce errors.

Since the dummy SQL utilizes views that encompass more comprehensive fields than those in tables, it is shorter in length than the final SQL. As a result, fewer tokens are needed to generate the dummy SQL, thereby decreasing the time needed for the LLM's output.

### 4.5 SQL Reconstruction module

The SQL reconstruction module transforms the dummy SQL into the final SQL using Q and the relevant mapping rules, as shown in Figure 1(C). The prompt template is illustrated in Figure 5 of the Appendix. To prevent interference from irrelevant tables in the database, schema linking is implemented using the table mapping rules from the view creation syntax. For example, in Figure 1(B), the view utilized by the dummy SQL is v_superhero. Based on the table mappings for v_superhero in Figure 1(C), the process selects the gender table, along with the colour and publisher tables. The output of this module is executable SQL designed for querying the actual database.

## 5. Experiments

### 5.1 Experiment Settings

**Datasets:** This section evaluates the performance of V-SQL on the Bird datasets. The Bird datasets is a challenging, large-scale, cross-domain text-to-SQL datasets. Follow previous works[2], a portion of the development set is split to create the test set, as it has not been released. This test set contains 11 databases and 500 pairs of text-to-SQL data. The distribution of SQL difficulty levels and the average number of joins in the test set is presented in Table 2. It is evident that as the difficulty of the questions increases, the average number of join operation also increases.

Table 2

SQL Statistics in the Bird Development Set

|  | SIM. | MOD. | CHALL. |
|---|---|---|---|
| number | 148 | 250 | 102 |
| Average number of join operation | 0.75 | 1.16 | 1.35 |
| The average number of tables per SQL query | 2.3 | 2.62 | 2.76 |
| The category of table combinations | 50 | 86 | 54 |

**Metrics:** Consistent with previous works[2], the evaluation metric is EX (Execution Accuracy). EX measures whether the predicted SQL is valid and whether its execution results match those of the ground truth SQL.

**Comparison Methods:** V-SQL is compared with three prominent ICL-based methods: DIN-SQL[5], DAIL-SQL[6] and TA-SQL[2]. To ensure fairness, none of the models utilized self-consistency or remodification mechanisms.

**Implementation Details:** Our method is implemented using closed-source LLMs, including ChatGPT (gpt-3.5-turbo)[7], GPT-4 (gpt-4-32k)[8], and GPT-4 Turbo. The model temperature is set to 0, and the top-p is set to 1. The maximum number of tokens for all models is limited to 800.

### 5.2 Main Results

As presented in Table 3, the test results on GPT-4, GPT-4-turbo, and GPT-3.5-turbo indicate that the performance of V-SQL is comparable to that of TA-SQL. Specifically, when using GPT-4-turbo and GPT-3.5-turbo, the overall EX metric of V-SQL surpasses TA-SQL by 2.64% and 1.60%, respectively. In contrast, when using GPT-4, TA-SQL outperforms V-SQL in the total EX metric. We speculate that this phenomenon is related to TA-SQL directly extracting tables and columns from dummy SQL for subsequent steps. As model capabilities improve, schema linking directly based on the model capabilities yields better results.

**Table 3**

Execution Accuracy (EX) (%) on Bird mini-dev datasets

| Model | METHOD | SIM. | MOD. | CHALL. | TOTAL |
|---|---|---|---|---|---|
| GPT4 | DIN-SQL | 63.51 | 43.60 | 32.35 | 47.2 |
| | DAIL-SQL | 66.89 | 44.4 | 34.31 | 49.0 |
| | TA-SQL | 62.83 | **60.4** | 36.11 | **56.17** |
| | V-SQL(Ours) | **62.84** | 59.6 | **37.25** | 56.0 |
| GPT-4-turbo | DIN-SQL | 58.11 | 47.6 | 33.3 | 47.8 |
| | DAIL-SQL | 60.13 | 48.8 | 32.35 | 48.8 |
| | TA-SQL | 62.9 | 48.4 | 36.11 | 50.16 |
| | V-SQLG(Ours) | **62.16** | **51.6** | **40.19** | **52.8** |
| GPT-35-turbo | DIN-SQL | 49.32 | 39.6 | 23.52 | 39.2 |
| | DAIL-SQL | 52.03 | 43.6 | 23.52 | 42.0 |
| | TA-SQL | 56.75 | 42.8 | 27.45 | 43.8 |
| | V-SQLG(Ours) | **58.11** | **43.6** | **31.37** | **45.4** |

**4.3 Fine-grained Case Study**

To illustrate the effectiveness of V-SQL intuitively, one of its generated results in the Bird datasets is analyzed as shown in Table 4. The superhero database is included in the Bird datasets. The superhero table, one of the tables in the superhero database, includes the foreign keys eye_colour_id, hair_colour_id, gender_id, race_id, publisher_id, and alignment_id. According to the table mapping rules in Table 4, the information relative to these foreign keys is merged into a view named v_superhero.

The output of the first stage is dummy SQL, which is significantly shorter than both the gold SQL and the final SQL. The dummy SQL will be reconstructed according to the table mapping rule in the next stage.

The output of the second stage is final SQL, which does not contain any view. The execution result of the final SQL is the same as the gold SQL.

## 6. Discussion

Due to the sensitivity of LLMs to prompts, the representation of database schemas in ICL-based text-to-SQL directly impacts the final results. Therefore, the effectiveness of the view-based schema strategy proposed in this paper is significantly influenced by the structure and description of the views. In practice, views are often created based on frequently used join queries. Hence, in this experiment, we referred to the join operations in the ground truth SQL from the dataset and manually designed the view creation syntax.

We experimented with various view design strategies and found that creating views for all join operations appearing in the ground truth SQL does not necessarily yield the best results. As the number of views increases, it becomes more challenging to retrieve the correct view description based on the user's query.

## 7. CONCLUSION

This paper proposes a view-based schema strategy that addresses the challenge of database schemas being difficult for LLMs to comprehend in text-to-SQL task. The strategy utilizes the view syntax of databases to eliminate the complexity introduced by foreign keys and map the related fields distributed across multiple tables into a single view. Furthermore, a two-stage text-to-SQL framework, named V-SQL, is introduced. V-SQL generates concise dummy SQL during the first stage. In the second stage, it reconstructs the dummy SQL into final SQL. Experimental results on the Bird datasets show that V-SQL achieves performance comparable to the current state-of-the-art models.

Currently, the design of table mappings requires manual effort. The automatic generation of the most suitable table mappings using LLMs remains a topic for further exploration.

**Table 4:**

An example of SQL Generated by V-SQL in the Bird Datasets.

| DB name | Superhero |
|---|---|
| Question | Please list the superhero names of all the superheroes that have blue eyes |
| Gold SQL | select \`T1\`.\`superhero_name\` <br> from \`superhero\` as \`T1\` <br> inner join \`colour\` as \`T2\` on \`T1\`.\`eye_colour_id\` = \`T2\`.\`id\` <br> inner join \`colour\` as \`T3\` on \`T1\`.\`hair_colour_id\` = \`T3\`.\`id\` <br> where \`T2\`.\`colour\` = 'Blue' and \`T3\`.\`colour\` = 'Blond' |
| Table Mapping Rule | create view v_superhero as select \`superhero\`.\`id\`, \`superhero\`.\`superhero_name\` as \`superhero_name\`, \`superhero\`.\`full_name\`, \`gender\`.\`gender\`, \`eye_colour\`.\`colour\` as \`eye_colour\`, \`hair_colour\`.\`colour\` as \`hair_colour\`, \`skin_colour\`.\`colour\` as \`skin_colour\`, \`race\`.\`race\`, \`publisher\`.\`publisher_name\`, \`alignment\`.\`alignment\`, \`superhero\`.\`height_cm\`, \`superhero\`.\`weight_kg\` <br> from \`superhero\` <br> left join \`gender\` on \`superhero\`.\`gender_id\` = \`gender\`.\`id\` <br> left join \`colour\` eye_colour on \`superhero\`.\`eye_colour_id\` = \`eye_colour\`.\`id\` <br> left join \`colour\` hair_colour on \`superhero\`.\`hair_colour_id\` = \`hair_colour\`.\`id\` <br> left join \`colour\` skin_colour on \`superhero\`.\`skin_colour_id\` = \`skin_colour\`.\`id\` <br> left join \`race\` on \`superhero\`.\`race_id\` = \`race\`.\`id\` <br> left join \`publisher\` on \`superhero\`.\`publisher_id\` = \`publisher\`.\`id\` <br> left join \`alignment\` on \`alignment\`.\`id\` = \`superhero\`.\`alignment_id\` |
| The output of the first stage | select \`v_superhero\`.\`superhero_name\` <br> from \`v_superhero\` <br> where \`v_superhero\`.\`eye_colour\` = 'Blue' and \`v_superhero\`.\`hair_colour\` = 'Blond' |
| The output of the second stage | select \`superhero\`.\`superhero_name\` <br> from \`superhero\` <br> inner join \`colour\` eye_colour ON \`superhero\`.\`eye_colour_id\` = \`eye_colour\`.\`id\` <br> inner join \`colour\` hair_colour ON \`superhero\`.\`hair_colour_id\` = \`hair_colour\`.\`id\` <br> where \`eye_colour\`.\`colour\` = 'Blue' and \`hair_colour\`.\`colour\` = 'Blond |

# Appendix A

**Figure 2**

Prompt template for view creation in the table mapping module

```
# task
You are a professional database administrator. I need to create views based on "db schema" to eliminate foreign keys as much as possible. Please tell me the SQL without explanation.

notice:
1. All views start with "v_"
2. Don't use alias
3. View don't select other view, use original table only.
4. try to merge relevant tables in order to reduce the frequency of using join

# db schema
{db_description}

# format example
```sql
your sql
```
```

**Figure 3**

An example of view creation in the table mapping module. After the table undergoes the table mapping operation, fields like gender_id, eye_colour_id, hair_color_id, skin_colour_id, race_id, publisher_id, and alignment_id are mapped to their specific values through join operations. This reduces the reliance on join operations in subsequent queries.

```sql
create view v_superhero as
select
   `superhero`.`id`,
   `superhero`.`superhero_name` as `superhero_name`,
   `superhero`.`full_name`,
   `gender`.`gender`,
   `eye_colour`.`colour` as `eye_colour`,
   `hair_colour`.`colour` as `hair_colour`,
   `skin_colour`.`colour` as `skin_colour`,
   `race`.`race`,
   `publisher`.`publisher_name`,
   `alignment`.`alignment`,
   `superhero`.`height_cm`,
   `superhero`.`weight_kg`
from
   `superhero`
left join `gender` on `superhero`.`gender_id` = `gender`.`id`
left join `colour` eye_colour on `superhero`.`eye_colour_id` = `eye_colour`.`id`
left join `colour` hair_colour on `superhero`.`hair_colour_id` = `hair_colour`.`id`
left join `colour` skin_colour on `superhero`.`skin_colour_id` = `skin_colour`.`id`
left join `race` on `superhero`.`race_id` = `race`.`id`
left join `publisher` on `superhero`.`publisher_id` = `publisher`.`id`
left join `alignment` on `alignment`.`id` = `superhero`.`alignment_id`
```

**Figure 4:**

The prompt used by the table mapping module in the first stage of V-SQL. In this prompt, varname db_schema represents the description of tables related to the query, retrieved from a database.

```
# target
You are an experienced database administrator, please answer "question" by sql with no explanation.
notice:
1. Do not alias the output fields
2. must avoid ambiguous column name by using table name and column name in the SQL statement. For example, use `table_name.column_name` instead of just `column_name`.
3. refer to "relevant db schema"

# relevant db schema
{relevant_db_schema}

# question
{query}
```

**Figure 5**

The prompt of the SQL reconstruction module in the second stage of V-SQL.

```
# target
According to the "views schemas", "view query" and "relevant table schemas", restore the sql in "view query" to one used the original table query without explanation
notice:
1. SQL should contain "join" as little as possible.
2. table in "views schemas" should not be contained in output.
3. comply with the syntax of {db_type}.
4. must avoid ambiguous column name by using table name and column name in the SQL statement. For example, use `table_name.column_name` instead of just `column_name`.

# view schemas
{view_schemas}

# view query
Question: {query}
```sql
{dummy_sql}
```

# relevant table schemas
{table_schemas}

# foramt
```sql
{sql}
```
```

**Table 6**

The superhero table in the Superhero Database from the Bird Datasets (fields name in red font indicate foreign keys)

| original_column_name | column_name | column_description | data_format |
|---|---|---|---|
| id | id | the unique identifier of the superhero | integer |
| superhero_name | superhero name | the name of the superhero | text |
| full_name | full name | the full name of the superhero | text |
| gender_id | gender id | the id of the superhero's gender | integer |
| eye_colour_id | eye colour id | the id of the superhero's eye color | integer |
| hair_colour_id | hair colour id | the id of the superhero's hair color | integer |
| skin_colour_id | skin colour id | the id of the superhero's skin color | integer |
| race_id | race id | the id of the superhero's race | integer |
| publisher_id | publisher id | the id of the publisher | integer |
| alignment_id | alignment id | the id of the superhero's alignment | integer |
| height_cm | height cm | the height of the superhero | integer |
| weight_kg | weight kg | the weight of the superhero | integer |